\documentclass[11pt,reqno]{amsart}
\usepackage{epsfig}
\usepackage{graphicx}
\usepackage[numbers,sort&compress]{natbib}
\begin{document}
\baselineskip=0.6cm 
\theoremstyle{plain}
\title{Infinitely many nonlocal symmetries and conservation laws for  the (1+1)-dimensional Sine-Gordon equation}
\author{Xiao-yan Tang$^{1}$, Zu-feng Liang$^{2}$ and Sen-yue Lou$^{3,4}$}
\dedicatory{$^1$Department of Physics and Astronomy, Shanghai Jiao Tong University, Shanghai 200240, China\\
$^2$ Department of Physics, Hangzhou Normal University, Hangzhou 310036, China\\
$^3$ Shanghai Key Laboratory of Trustworthy Computing, East China Normal University, Shanghai 200062, China\\
$^4$ Faculty of Science, Ningbo University, Ningbo, 315211, China}

\date{13 August 2013}
 
\begin{abstract}
Infinitely many nonlocal symmetries and  conservation laws of the (1+1)-dimensional Sine-Gordon (SG) equation are derived in terms of its B\"acklund transformation (BT). Some special nonlocal symmetries and nonlocal conservation laws are obtained from the linearized equations of the SG equation and its BT.  Furthermore, one can derive infinitely many nonlocal symmetries from a known nonlocal symmetry, but also infinitely many nonlocal conservation laws from a known nonlocal conservation law.  In addition, infinitely many local and nonlocal conservation laws can be directed generated from BT through the parameter expansion procedure.
\end{abstract}

\maketitle




\section{Introduction}

Symmetries and conservation laws belong to the central studies of nonlinear evolutional  equations. Especially,  one nonlinear partial differential equation (NPDE) is believed to be integral in the sense that it possesses infinite number of symmetries or conservation laws. Besides, one can construct one or more conservation laws from one known symmetry, but almost all the conservation laws of PDEs may not have physical interpretations except for several well known cases, such as the invariance of the spatial transformation ensures the conservation of momentum and the invariance of the temporal transformation guarantees the energy conservation. 

There are many effective methods to find symmetries and conservation laws of a PDE. Generally, one can use the classical Lie group approach, or equivalently, the symmetry approach, to discover classical Lie symmetries, while nonclassical Lie symmetries can be obtained by means of the nonclassical Lie group approach involving the prolongation structures \cite{cllie,noncllie}. Both of them can be recovered by the direct method \cite{direct1,direct2,direct3} without using any group theory. However, these methods cannot be applied directly to find nonlocal symmetries and even higher order local symmetries. Up to now, various nonlocal symmetries have been investigated such as the potential symmetries which can be obtained by applying the Lie group approach to the potential form of the given system.  Another type of well studied nonlocal symmetries is the so-called eigenfunction symmetries \cite{eigensym1,eigensym2,eigensym3,eigensym4}. Lately, eigenfunction symmetries of the Korteweg-de Vries (KdV) equation  related to the Darboux transformations have been used to obtain explicit interesting interacting wave solutions \cite{hulouchen}. Nonlocal symmetries related to B\"acklund transformations (BT) have also been studied \cite{louhuchen}. Recently, the residual symmetry \cite{superkdv} is proposed and thus obtained for the supersymmetric KdV equations through the truncated Painlev\'e expansion method and they are actually related to BT.

It is known that conservation laws have a close relation with symmetries, and moreover, have many important applications \cite{law}. In addition to the description of physical conserved quantities such as mass, energy, momentum as well as angular momentum, and the significance for investigating integrability as mentioned above, they can also be used in the analysis of stability and global behavior of solutions,  play an essential role in numerical methods, and so on. In order to construct conservation laws, different methods have been established, for instance, the celebrated Noether's theorem \cite{noether}, the conservation law multiplier approach \cite{multiplier}, and the characteristic method \cite{characteristic}. 

In this paper, we focus our attention on the nonlocal symmetries and nonlocal conservation laws related to BT. Take the (1+1)-dimensional Sine-Gordon (SG)  equation
\begin{equation} \label{sgu}
u_{xt}=m^2\sin u, 
\end{equation}
where $m$ is an arbitrary constant, as an concrete example. The sine-Gordon equation has various physical applications, for instance in relativistic field theory, Josephson junctions, mechanical transmission lines, and nonlinear optics. The symmetries, invariance group transformations, local and nonlocal conservation laws for the SG equation have already been studied \cite{sg1,sg2}. Here, in a different way, it is shown that one can obtain some novel special nonlocal symmetries with a nonlocal function and a parameter coming from BT of the given system. Thereafter, through the parameter expansion, new series of infinite number of nonlocal symmetries can be constructed from one special nonlocal symmetry. From the symmetry equations of the given SG equation and its BT, one can also obtain some parameter dependent  nonlocal conservation laws which could be related to the nonlocal symmetries. In the same manner, infinitely many nonlocal conservation laws can be produced from one nonlocal conservation law via the parameter expansion. On the other hand, it is clear that the compatibility condition of BT for the SG equation gives the SG equation itself, and thus it is demonstrated that infinite number of local and nonlocal conservation laws can also be obtained if one expands BT in terms of its parameter. 

It is well known that the SG equation \eqref{sgu} has the following auto-B\"acklund transformation\begin{eqnarray}
&&u_x-v_x=2\lambda m\sin\left(\frac u 2+\frac v 2\right),\label{bt1}\\
&&u_t+v_t=2\frac m {\lambda} \sin\left(\frac u 2-\frac v 2\right),\label{bt2}
\end{eqnarray}
which means that if $u$ is a solution of the SG equation \eqref{sgu}, then $v$ determined by Eqs. \eqref{bt1} and \eqref{bt2}  also satisfies the SG equation, namely,
\begin{eqnarray}
v_{xt}=m^2\sin v.  \label{sgv}
\end{eqnarray}

The paper is organized as follows. In section 2, the linearized differential equations, i. e., the symmetry equations, of Eqs. \eqref{sgu}-\eqref{sgv} are written down from which nonlocal symmetries of the SG equation \eqref{sgu} can be derived. In detail, three special solutions of the symmetry equations are explicitly written down, which are really the nonlocal symmetries of the SG equation. It is obvious that these special nonlocal symmetries are parameter dependent, therefore, infinitely many nonlocal symmetries can be generated from them through the parameter expansion, which is demonstrated in detail for one of the special nonlocal symmetries. In section 3, from solving the symmetry equations, three special nonlocal conservation laws related to BT are derived. It is quite natural and straightforward to find that they are related to the nonlocal symmetries.  As the nonlocal conservation laws are also parameter dependent, in the same way, infinitely many nonlocal conservation laws can be constructed from one nonlocal conservation law via the parameter expansion. Explicitly, one series of infinite number of nonlocal conservation laws are presented. In section 4, applying the same expansion directly to  BT \eqref{bt1}-\eqref{bt2}, it is shown that not only infinitely many nonlocal conservation laws but also infinitely many local conservation laws can be obtained for the SG equation. The last section is devoted to summary and discussions.

\section{Nonlocal symmetries related to BT}

Symmetries of a PDE are actually the solutions of the linearized equation obtained from the requirement that the PDE is form invariant under an infinitesimal transformation. Assuming that Eqs. \eqref{sgu}-\eqref{sgv} are form invariant under the following transformation
\begin{eqnarray}
u\rightarrow u+\epsilon \sigma^u,\quad  v\rightarrow v+\epsilon \sigma^v,\quad  \lambda\rightarrow \lambda+\epsilon\delta,
\end{eqnarray}
with $\epsilon$ being an infinitesimal parameter, we readily obtain the linearized system that symmetries $\sigma^u$ and $ \sigma^v$ should satisfy
\begin{eqnarray}
&&\sigma^u_{xt}-m^2\cos(u)\sigma^u=0,\label{rsymu}\\
&&\sigma^v_{xt}-m^2\cos(v) \sigma^v=0,\label{rsymv}\\
&&\sigma^u_x-\sigma^v_x-2\delta m \sin\left(\frac u 2+\frac v 2\right)-\lambda m\cos\left(\frac u 2+\frac v 2\right)(\sigma^u+\sigma^v)=0,\label{rsymbt1}\\
&&\sigma^u_t+\sigma^v_t+2\frac{m\delta}{\lambda^2} \sin\left(\frac u 2-\frac v 2\right)-\frac{m}{\lambda}\cos\left(\frac u 2-\frac v 2\right)(\sigma^u-\sigma^v)=0.\label{rsymbt2}
\end{eqnarray}

It is  remarkable that $\sigma^u$ obtained from Eqs.  \eqref{rsymu}-\eqref{rsymbt2} depends on the function $v$, which is related to the function $u$ through BT \eqref{bt1}-\eqref{bt2}, therefore, the solution $\sigma^u$ is actually a nonlocal symmetry of the SG equation \eqref{sgu}. Evidently,  two undetermined symmetries $\sigma^u$ and $ \sigma^v$ should satisfy four linear differential equations \eqref{rsymu}-\eqref{rsymbt2}, so unfortunately, it is still rather difficult to obtain their general solutions. However, it is possible to write down some special solutions.  In the following we present three possible special solutions of Eqs. \eqref{rsymu}-\eqref{rsymbt2}. 

{\bf Solution I.} The first special nonlocal symmetry is obtained without difficulty by assuming  $\delta=0$ and $\sigma^v=0$. In this case, the nonlocal symmetry  $\sigma^u$ reads
\begin{eqnarray}
\sigma_1^u=e^{\lambda m p},\label{nonsym1}
\end{eqnarray}
where $p$ is determined by
\begin{eqnarray}
p_x=\cos\left(\frac u 2 +\frac v 2\right),\quad p_t=\frac 1 {\lambda^2}\cos
\left(\frac u 2 -\frac v 2\right).\label{pxt1}
\end{eqnarray}
It is noted that the consistent condition $p_{xt}=p_{tx}$ is satisfied by using BT \eqref{bt1}-\eqref{bt2}.

{\bf Solution II.} If still assuming that $\sigma^v=0$ and then requiring $\delta=1/2m$,  we can have the second nonlocal symmetry as
\begin{eqnarray}
\sigma_2^u=qe^{\lambda m p},
\end{eqnarray}
where $p$ and $q$ are given by
\begin{eqnarray}
p_x=\cos\left(\frac u 2 +\frac v 2\right),\quad p_t=\frac 1 {\lambda^2}\frac{\cos(u)+\cos(v)}{\cos
\left(\frac u 2-\frac v 2\right)},
\end{eqnarray}
and
\begin{eqnarray}
q_x=e^{-\lambda mp}\sin\left(\frac u 2 +\frac v 2
\right),\quad q_t=\frac 1 {\lambda^2}\frac{\sin(u)-\sin(v)}{\cos\left(\frac u 2+\frac v 2
\right)}e^{-\lambda mp},
\end{eqnarray}
respectively.
It is remarkable again that $p_{xt}=p_{tx}$ and $q_{xt}=q_{tx}$ are satisfied identically by utilizing BT \eqref{bt1}-\eqref{bt2}.

{\bf Solution III.} It is also possible to obtain some special solutions when the symmetry $\sigma^v$ is nonzero. From the SG equation \eqref{sgv}, one can easily verify that  $\sigma^v=v_x$ is a solution of Eq. \eqref{rsymv}. Then solving the remaining three symmetry equations in the case of $\delta=0$, we have the following special nonlocal symmetry
\begin{eqnarray}
\sigma_3^u=v_x+2m\lambda fe^{\lambda m p},
\end{eqnarray}
where $f$ is determined by
\begin{eqnarray}
f_x=v_x\cos
\left(\frac u 2 +\frac v 2\right)e^{-\lambda mp}, \quad f_t=-\lambda m v_{xt}e^{-\lambda mp},\end{eqnarray}
and $p$ satisfies Eq. \eqref{pxt1}. It is readily proved that the consistent condition $f_{xt}=f_{tx}$ is satisfied by virtue of the SG equation \eqref{sgv} and BT \eqref{bt1}-\eqref{bt2} .

It is obvious that the above three special nonlocal symmetries are parameter dependent, therefore, one can go further to obtain infinitely many nonlocal symmetries through the expansions of the special nonlocal symmetries around the neighborhood of the spectrum parameter $\lambda$ coming from BT. Here we just present a series of infinite number of nonlocal symmetries from the first special nonlocal symmetry \eqref{nonsym1}. In the same way, another two series of infinitely many nonlocal symmetries can be produced. 

First, we expand the function $p$ as
\begin{eqnarray}
p=\sum_{i=0}^{\infty} p_i\delta^i, \label{pexpansion}
\end{eqnarray}
where the expansion coefficients $p_i$ are functions of $x$ and $t$, and $\delta$ is an arbitrary expansion constant. Inserting the expansion \eqref{pexpansion} into the nonlocal symmetry \eqref{nonsym1} with $\lambda$ replaced by $\lambda+\delta$, we have
\begin{eqnarray}
\sigma_1^u=\exp\left({(\lambda+\delta)m\sum_{i=0}^{\infty} p_i\delta^i}\right)\equiv \sum_{i=0}^{\infty} \sigma_{1i}^u\delta^i,\label{sigmai}
\end{eqnarray}
where  $\sigma_{1i}^u$ can be computed by the formula
\begin{eqnarray}
\sigma_{1i}^u=\frac{d^{n}\sigma_1^u}{d\delta^n}|_{\delta=0}.\label{defsigmai}
\end{eqnarray}

It is remarkable that all the expansion coefficients $\sigma_{1i}^u$ are nonlocal symmetries satisfying the symmetry equation \eqref{rsymu}. Therefore, from the special nonlocal symmetry \eqref{nonsym1}, infinitely many nonlocal symmetries $\sigma_{1i}^u$ are generated, which obviously involve the functions $p_i$ determined by Eq. \eqref{pxt1} with the similar parameter expansion. Furthermore, because the functions $u$ and $v$ in Eq. \eqref{pxt1} are connected with each other through  BT \eqref{bt1}-\eqref{bt2}, we then also need to make the similar parameter expansion for $v$. Although it is difficult to write down the general explicit expression for infinitely many nonlocal symmetries $\sigma_{1i}^u$, it is quite easy and straightforward to write down the explicit forms one by one from the general expansion formula.  Below are the explicit descriptions for the first five nonlocal symmetries $\sigma_{1i}^u$ for $i=0,1,2,3,4$,
\begin{eqnarray}
\sigma^u_{10}&=&e^{m\lambda p_0},\label{resigma0}\\
\sigma^u_{11}&=&m(p_0+\lambda p_1)e^{m\lambda p_0},\label{resigma1}\\
\sigma^u_{12}&=&m\left(\frac 1 2
m(p_0+\lambda p_1)^2+(p_1+\lambda p_2)\right)e^{m\lambda p_0},\label{resigma2}\\
\sigma^u_{13}&=&m\left(\frac 1 6
m^2(p_0+\lambda p_1)^3+m(p_0+\lambda p_1)(p_1+\lambda p_2)+(p_2+\lambda p_3)\right)e^{m\lambda p_0},\label{resigma3}\\
\sigma^u_{14}&=&m\left[\frac 1 {24}
m^3(p_0+\lambda p_1)^4+\frac 1 2m^2(p_0+\lambda p_1)^2(p_1+\lambda p_2)\right.\nonumber\\
&&\left.+m\left(\frac 12(p_1+\lambda p_2)^2+(p_0+\lambda p_1)(p_2+\lambda p_3)\right)+(p_3+\lambda p_4)\right]e^{m\lambda p_0},\label{resigma4}
\end{eqnarray}
where $p_i~(i=0,1,2,3,4)$ are determined  by 
\begin{eqnarray}
&& p_{0x}=\cos\left(\frac u 2+\frac {v_0} 2\right),\label{p0x}\\
 && p_{0t}= \frac 1 {\lambda^2} \cos\left(\frac u 2-\frac {v_0} 2\right),\label{p0t}
\end{eqnarray}
\begin{eqnarray}
&&p_{1x}=-\frac 1 2v_1\sin\left(\frac u 2+\frac {v_0} 2\right),\label{p1x}\\
&&p_{1t}= -\frac 2 {\lambda^3} \cos\left(\frac u 2-\frac {v_0} 2\right)+\frac 1 {2\lambda^2}v_1 \sin\left(\frac u 2-\frac {v_0} 2\right),\label{p1t}
\end{eqnarray}
\begin{eqnarray}
&&p_{2x}=-\frac 1 8v_1^2\cos\left(\frac u 2+\frac {v_0} 2\right)-\frac 1 2v_2\sin\left(\frac u 2+\frac {v_0} 2\right),\label{p2x}\\
&&p_{2t}= \left(\frac{3}{\lambda^4}-\frac{v_1^2}{8\lambda^2}\right) \cos\left(\frac u 2-\frac {v_0} 2\right)-\left(\frac{v_1}{\lambda^3}-\frac{v_2}{2\lambda^2}\right) \sin\left(\frac u 2-\frac {v_0} 2\right),\label{p2t}
\end{eqnarray}
\begin{eqnarray}
&&p_{3x}=-\frac 1 4v_1v_2\cos\left(\frac u 2+\frac {v_0} 2\right)+\frac 1 {48}(v_1^3-24v_3)\sin\left(\frac u 2+\frac {v_0} 2\right),\label{p3x}\\
&&p_{3t}= \left(- \frac {v_1v_2} {4\lambda^2}+\frac{v_1^2}{4\lambda^3}+\frac 4 {\lambda^5} \right) \cos\left(\frac u 2-\frac {v_0} 2\right)+\left(\frac{24_3-v_1^3}{48\lambda^2}-\frac{v_2}{\lambda^3}+\frac{3v_1}{2\lambda^4}\right) \sin\left(\frac u 2-\frac {v_0} 2\right),\label{p3t}
\end{eqnarray}
and
\begin{eqnarray}
&&p_{4x}=\left(\frac{v_1^4}{384}-\frac{v_2^2}8-\frac{v_1v_3}4\right)\cos\left(\frac u 2+\frac {v_0} 2\right)+\left(\frac{v_1^2v_2}{16}-\frac{v_4}2\right)\sin\left(\frac u 2+\frac {v_0} 2\right),\label{p4x}\\
&&p_{4t}= \left(\frac{v_1^4}{384\lambda^2}-\frac{v_2^2}{8\lambda^2}-\frac{v_1v_3}{4\lambda^2}+\frac{v_1v_2}{2\lambda^3}-\frac{3v_1^2}{8\lambda^4}+\frac{5}{\lambda^6}\right) \cos\left(\frac u 2-\frac {v_0} 2\right)\nonumber\\
&&\qquad+\left(-\frac{v_1^2v_2}{16\lambda^2}+\frac{v_4}{2\lambda^2}+\frac{v_1^3-24v_3}{24\lambda^3}+\frac{3v_2}{2\lambda^4}-\frac{2v_2}{\lambda^5}\right) \sin\left(\frac u 2-\frac {v_0} 2\right),\label{p4t}
\end{eqnarray}
respectively, obtained by substituting Eq. \eqref{pexpansion} and  the expansion of $v$
\begin{eqnarray}
v=\sum_{i=0}^{\infty} v_i\delta^i, \label{vexpansion}
\end{eqnarray}
into Eq. \eqref{pxt1} with  $\lambda$ replaced by $\lambda+\delta$, and then equating zero all the coefficients of different powers of $\delta$.  The $v_i~(i=0,1,2,3,4)$ in Eqs. \eqref{p0x}-\eqref{p4t} are determined by the system of
\begin{eqnarray}
&&v_{0x}=u_x-2m\lambda\sin\left(\frac u 2+\frac {v_0} 2\right),\label{v0x}\\
&&v_{0t}=-u_t+\frac{2m}{\lambda}\sin\left(\frac u 2-\frac {v_0} 2 \right),\label{v0t}
\end{eqnarray}
\begin{eqnarray}
&&v_{1x}=-m \lambda v_1\cos\left(\frac u 2+\frac {v_0} 2\right)-2m\sin\left(\frac u 2+\frac {v_0} 2\right),\label{v1x}\\
&&v_{1t}=-\frac m {\lambda}v_1\cos\left(\frac u 2-\frac {v_0} 2\right)-\frac {2m} {\lambda^2}\sin\left(\frac u 2-\frac {v_0} 2\right).\label{v1t}
\end{eqnarray}
\begin{eqnarray}
&&v_{2x}=-m (v_1+\lambda v_2)\cos\left(\frac u 2+\frac {v_0} 2\right)+\frac 1 4m\lambda v_1^2\sin\left(\frac u 2+\frac {v_0} 2\right),\label{v2x}\\
&&v_{2t}=m\left(\frac{v_1}{\lambda^2}-\frac{v_2}{\lambda}\right)\cos\left(\frac u 2-\frac {v_0} 2\right)+m\left(\frac{2}{\lambda^3}-\frac{v_1^2}{4\lambda}\right)\sin\left(\frac u 2-\frac {v_0} 2\right),\label{v2t}
\end{eqnarray}
\begin{eqnarray}
&&v_{3x}=\frac m {24} (\lambda (v_1^3-24v_3)-24v_2)\cos\left(\frac u 2+\frac {v_0} 2\right)+\frac 1 4mv_1(v_1+2\lambda v_2)\sin\left(\frac u 2+\frac {v_0} 2\right),\label{v3x}\\
&&v_{3t}=m\left(\frac{v_1^3-24v_3}{24\lambda}+\frac{v_2}{\lambda^2}-\frac{v_1}{\lambda^3}
\right)\cos\left(\frac u 2-\frac {v_0} 2\right)-m\left(\frac{v_2v_1}{2\lambda}-\frac{v_1^2}{4\lambda^2}+\frac{2}{\lambda^4}
\right)\sin\left(\frac u 2-\frac {v_0} 2\right),\label{v3t}
\end{eqnarray}
and
\begin{eqnarray}
&&v_{4x}=\frac{m}{24}(3\lambda(v_1^2v_2-8v_4)+(v_1^3-24v_3))\cos\left(\frac u 2+\frac {v_0} 2\right)\nonumber\\
&&\qquad\quad+\frac m {192}(\lambda(48v_2^2+96v_3v_1-v_1^4)+96v_1v_2)\sin\left(\frac u 2+\frac {v_0} 2\right),\label{v4x}\\
&&v_{4t}=m\left(\frac{v_2v_1^2-8v_4}{8\lambda}-\frac{v_1^3-24v_3}{24\lambda^2}-\frac{v_2}{\lambda^3}+\frac{v_1}{\lambda^4}\right)\cos\left(\frac u 2-\frac {v_0} 2 \right)\nonumber\\
&&\qquad\quad+m\left(\frac{v_1^4-96v_1v_3-48v_2^2}{192\lambda}+\frac{v_1v_2}{2\lambda^2}-
\frac{v_1^2}{4\lambda^3}+\frac{2}{\lambda^5}\right)\sin\left(\frac u 2-\frac {v_0} 2\right),\label{v4t}
\end{eqnarray}
respectively, which are obtained similarly by inserting Eq. \eqref{vexpansion} into BT \eqref{bt1}-\eqref{bt2} with $\lambda$ replaced by $\lambda+\delta$, and then equating zero all the coefficients of different orders of $\delta$.

It is easy to check that the consistent conditions $v_{ixt}=v_{itx}~(i=0,1,2,3,4)$ are all satisfied identically with the help of the SG equation \eqref{sgu}, while the consistent conditions $p_{ixt}=p_{itx}~(i=0,1,2,3,4)$ are all satisfied identically with $v_i~(i=0,1,2,3,4)$ determined by Eqs. \eqref{v0x}-\eqref{v4t}.  In addition, it is not difficult to verify that Eqs. \eqref{resigma0}-\eqref{resigma4} are all nonlocal symmetries of the SG equation \eqref{sgu} by the direct substitution of them with Eqs. \eqref{p0x}-\eqref{p4t} and \eqref{v0x}-\eqref{v4t} into the corresponding symmetry equation \eqref{rsymu}. 

\section{Nonlocal conservation laws related to the nonlocal symmetries}

The essence for a conservation law of a PDE is to find a divergence expression holding for all solutions of the given PDE. In detail, for our case, if there are two differentiable functions $\rho$ and  $\theta$ satisfy the identity
\begin{eqnarray}
\frac{\partial}{\partial t} \rho(x,t)+\frac{\partial}{\partial x}J(x,t)=0,\label{CLaw}
\end{eqnarray}
for any solution $u$ of the SG equation \eqref{sgu}, then this identity is called the conservation law of the SG equation, and the functions $\rho$ and $J$ are called the conserved density and the conserved flux, correspondingly. 

It is found that when solving the symmetry equations \eqref{rsymbt1} and \eqref{rsymbt2}, some auxiliary functions are introduced and the relations between these functions can rightly yield this kind of divergence expression \eqref{CLaw} holding on the solution of the SG equation. Therefore,  without using any known methods,  we can obtain several new special nonlocal conservation laws, as is shown in the following.

Now directly solving Eqs. \eqref{rsymbt1} and \eqref{rsymbt2}, namely, the symmetry equations of the BT, yields
\begin{eqnarray}
\sigma^u=2m e^{m\lambda p}q\delta+2 m e^{m\lambda p}\lambda\int p_x\sigma^ve^{-m\lambda p}{\rm d} x+\sigma^v+e^{m\lambda p}C,\label{sigmax}
\end{eqnarray}
and
\begin{eqnarray}
\sigma^u=-\frac{2ms}{\lambda^2}e^{\frac{mr}{\lambda}}\delta-\frac{2 m}{\lambda} e^{\frac{mr}{\lambda}}\int r_t\sigma^ve^{-\frac{mr}{\lambda}}{\rm d} t-\sigma^v+e^{\frac{mr}{\lambda}}C,\label{sigmat}
\end{eqnarray}
respectively, where $p,~q,~r,~s$ satisfy
\begin{eqnarray}
&& p_x=\cos\left(\frac u 2+\frac v 2 \right),\label{clpx}\\
&& q_x= \sin\left(\frac u 2 +\frac v 2\right)e^{m\lambda p},
\label{clqx}\\
&& r_t=\cos\left(\frac u 2-\frac v 2 \right),\label{clrt}\\
&& s_t=\sin\left(\frac u 2-\frac v 2\right)e^{-\frac{mr}{\lambda}},\label{clst}
\end{eqnarray}
and both integration functions are simplified to be a same constant $C$. The equivalence of Eqs. \eqref{sigmax} and \eqref{sigmat} and the arbitrariness of the constants $\delta$ and $C$ allow us to derive three equations
\begin{eqnarray}
r-\lambda^2 p=0,\label{claw1}
\end{eqnarray}
\begin{eqnarray}
s+\lambda^2 q=0,\label{claw2}
\end{eqnarray}
and
\begin{eqnarray}
\int p_x\sigma^ve^{-m\lambda p}{\rm d}x+\int p_t\sigma^ve^{-m\lambda p}{\rm d}t+\frac 1 {m\lambda} \sigma^v e^{-m\lambda p}=0.\label{claw3}
\end{eqnarray}
Hereafter, one can, without any difficulty, write down three kinds of nontrivial nonlocal conservation laws from the above three equations \eqref{claw1}-\eqref{claw3} with the help of Eqs. \eqref{clpx}-\eqref{clst}. The corresponding conserved density and flux are given by
\begin{eqnarray}
\rho_1=\lambda^2\cos\left(\frac u 2+\frac v 2\right),\quad J_1=\cos\left(\frac u 2-\frac v 2\right),\label{rJ1}
\end{eqnarray}
\begin{eqnarray}
\rho_2=\frac{\lambda^2}{\sigma_1^u}\sin\left(\frac u 2+\frac v 2 \right),\quad J_2=\frac{1}{\sigma_1^u},\label{rJ2}\cos\left(\frac u 2-\frac v 2 \right),
\end{eqnarray}
and
\begin{eqnarray}
\rho_3=\frac{\lambda\sigma^v_x}{m\sigma_1^u},\quad J_3=-\frac{\sigma^v}{\sigma_1^u}\cos\left(\frac u 2 -\frac v 2 \right),\label{rJ3}
\end{eqnarray}
respectively, where $\sigma_1^u$ is given by Eq. \eqref{nonsym1}, and $\sigma^v$ is the symmetry of the function $v$ determined by Eqs. \eqref{rsymu}-\eqref{rsymbt2}. It is evident that these nonlocal conservation laws are connected with the nonlocal symmetries. Similarly, the above nonlocal conservation laws are parameter dependent, therefore, one can construct infinitely many nonlocal conservation laws from any one of them via the parameter expansion procedure. 

Here we just present the result obtained from the nonlocal conservation law \eqref{rJ1}. Substituting Eq. \eqref{vexpansion} into the conserved density and flux \eqref{rJ1} with $\lambda$ replaced by $\lambda+\delta$, and then collecting the same orders of $\delta$, infinitely many nonlocal conservation laws $\partial_t\rho_{1i}(x,t)+\partial_xJ_{1i}(x,t)=0,~ i=0,1,2,...,$ are generated. The first five nonlocal conserved density are given explicitly as follows
\begin{eqnarray}
\rho_{10}=\lambda^2\cos\left(\frac u 2+\frac {v_0} 2\right),\label{rho10}
\end{eqnarray}
\begin{eqnarray}
\rho_{11}=2\lambda\cos\left(\frac u 2+\frac {v_0} 2 \right)
-\frac 1 2 \sin\left(\frac u 2+\frac {v_0} 2\right),\label{rho11}
\end{eqnarray}
\begin{eqnarray}
\rho_{12}=\left(1-\frac 1 8\lambda^2v_1^2\right)\cos\left(\frac u 2+\frac {v_0} 2 \right)
-\lambda\left(v_1+\frac 1 2\lambda v_2\right)\sin\left(\frac u 2+\frac {v_0} 2 \right),\label{rho12}
\end{eqnarray}
\begin{eqnarray}
\rho_{13}=\frac 1 4\lambda v_1(v_1+\lambda v_2)\cos\left(\frac u 2+\frac {v_0} 2\right)
+\left[\lambda\left(\frac 1 {48}v_1^3-\frac 1 2v_3\right)-\lambda v_2-\frac 1 2 v_1\right]\sin\left(\frac u 2+\frac {v_0} 2\right),\label{rho13}
\end{eqnarray}
\begin{eqnarray}
&&\rho_{14}=\left[\lambda^2\left(\frac 1{384}v_1^4-\frac 1 8 v_2^2-\frac 14v_1v_3\right)-\frac 12\lambda v_1v_2-\frac 18v_1^2\right]\cos\left(\frac u 2+\frac {v_0} 2\right)\nonumber\\
&&\qquad
+\left[\lambda^2\left(\frac 1 {16}v_1^2v_2-\frac 1 2v_4\right)+\lambda\left(\frac 1 {24}v_1^3-v_3\right)-\frac 1 2 v_2\right]\sin\left(\frac u 2+\frac {v_0} 2\right),\label{rho14}
\end{eqnarray}
and the corresponding nonlocal conserved flux read
\begin{eqnarray}
J_{10}=\cos\left(\frac u 2-\frac {v_0} 2\right),\label{J10}
\end{eqnarray}
\begin{eqnarray}
J_{11}=\frac 1 2 v_1\sin\left(\frac u 2-\frac {v_0} 2\right),\label{J11}
\end{eqnarray}
\begin{eqnarray}
J_{12}=-\frac 1 8 v_1^2\cos\left(\frac  u 2-\frac {v_0} 2\right)+\frac 1 2 v_2\sin\left(\frac u 2-\frac {v_0} 2\right),\label{J12}
\end{eqnarray}
\begin{eqnarray}
J_{13}=-\frac 1 4 v_1v_2\cos\left(\frac u 2-\frac {v_0} 2\right)-\left(\frac 1 {48} v_1^3-\frac 1 2 v_3\right)\sin\left(\frac u 2-\frac {v_0} 2\right),\label{J13}
\end{eqnarray}
\begin{eqnarray}
J_{14}=\left(\frac 1 {384} v_1^4-\frac 1 8 v_2^2-\frac 1 4 v_1v_3\right)\cos\left(\frac u 2-\frac {v_0} 2\right)-\left(\frac 1 {16} v_1^2v_2-\frac 1 2 v_4\right)\sin\left(\frac u 2-\frac {v_0} 2 \right),\label{J14}
\end{eqnarray}
respectively, where $v_i~(i=0,1,2,3,4)$ are determined by Eqs. \eqref{v0x}-\eqref{v4t}.

\section{Local and Nonlocal conservation laws related to BT}

It is clear that any two functions satisfying Eq. \eqref{CLaw} can be viewed as a conservation law. Therefore, one can naturally expect infinitely many conservation laws of the SG equation \eqref{sgu} using the compatibility condition of its BT \eqref{bt1}-\eqref{bt2} through the parameter expansion method. The BT has been expanded in section 2, as given by Eqs. \eqref{v0x}-\eqref{v4t}, one can then write down explicitly the corresponding conservation laws from the compatibility conditions $\partial_tv_{ix}=\partial_xv_{it},~i=0,1,2,3,4.$ The conserved density and conserved flux have the following form
\begin{eqnarray}
\rho_0=u_x-2m\lambda\sin\left(\frac u 2+\frac {v_0} 2\right),\label{nonlrho0}
\end{eqnarray}
\begin{eqnarray}
J_0=u_t-\frac {2m}{\lambda}\sin\left(\frac u 2-\frac {v_0} 2\right),
\end{eqnarray}

\begin{eqnarray}
\rho_1=-2m\sin\left(\frac u 2+\frac {v_0} 2\right)-mv_1\lambda\cos\left(\frac u 2+\frac {v_0} 2\right),
\end{eqnarray}
\begin{eqnarray}
J_1=\frac{2m}{\lambda^2}\sin\left(\frac u 2-\frac {v_0} 2\right)+\frac{mv_1}{\lambda}\cos\left(\frac u 2-\frac {v_0} 2\right),
\end{eqnarray}

\begin{eqnarray}
\rho_2=\frac 1 4m\lambda v_1^2\sin\left(\frac u 2+\frac {v_0} 2\right)-m(v_1+\lambda v_2)\cos\left(\frac u 2+\frac {v_0} 2\right),
\end{eqnarray}
\begin{eqnarray}
J_2=\frac{m(v_1^2\lambda^2-8)}{4\lambda^3}\sin\left(\frac u 2-\frac {v_0} 2\right)-\frac{m(v_1-v_2\lambda)}{\lambda^2}\cos\left(\frac u 2-\frac {v_0} 2\right),
\end{eqnarray}

\begin{eqnarray}
&&\rho_3=\frac 1 4 v_1m (v_1+2v_2\lambda)\sin\left(\frac u 2+\frac {v_0} 2\right)\nonumber\\
&&\qquad +\frac m {24}(v_1^3\lambda-24v_2-24\lambda v_3)\cos\left(\frac u 2+\frac {v_0} 2\right),
\end{eqnarray}
\begin{eqnarray}
&&J_3=\frac{m}{4\lambda^4}(8-v_1^2\lambda^2+2v_1v_2\lambda^3)\sin\left(\frac u 2-\frac {v_0} 2\right)\nonumber\\
&&\qquad +\frac{m}{24\lambda^3}(24v_1-v_1^3\lambda^2-24v_2\lambda+24v_3\lambda^2)\cos\left(\frac u 2-\frac {v_0} 2\right),
\end{eqnarray}

\begin{eqnarray}
&&\rho_4=-\frac m {192} (v_1^4\lambda-48v_2^2\lambda-96v_1v_2-96v_1v_3\lambda)\sin\left(\frac u 2+\frac {v_0} 2\right)\nonumber\\
&&\qquad +\frac m {24}(v_1^3+3\lambda v_1^2v_2-24v_3-24\lambda v_4)\cos\left(\frac u 2+\frac {v_0} 2\right),
\end{eqnarray}
and
\begin{eqnarray}
&& J_4=-\frac{m}{192\lambda^5}(384-48v_1^2\lambda^2+96v_1v_2\lambda^3-48v_2^2\lambda^4+v_1^4\lambda^4-96v_1v_3\lambda^4)\sin\left(\frac u 2-\frac {v_0} 2\right)\nonumber\\
&&\qquad +\frac{m}{24\lambda^4}(24v_2\lambda-24v_1-3v_1^2v_2\lambda^3+24v_4\lambda^3+v_1^3\lambda^2-24v_3\lambda^2)\cos\left(\frac u 2-\frac {v_0} 2\right),\label{nonlJ4}
\end{eqnarray}
respectively. 

It is remarkable that the conservation laws determined by Eqs. \eqref{nonlrho0}-\eqref{nonlJ4} are nonlocal conservation laws of the SG equation \eqref{sgu} derived with the help of its BT \eqref{bt1}-\eqref{bt2}. It is interesting that in addition to infinitely many nonlocal conservation laws, we can also construct infinitely many local conservation laws if the functions are expanded in the parameter $\lambda$ at the zero position. In this case, just substituting the expansion  \eqref{vexpansion} into Eqs. \eqref{bt1}-\eqref{bt2} with $\lambda$ replaced by $\delta$, and then setting zero the coefficients of different orders of $\delta$, a system of equations determining the expansion coefficient functions $v_i$ is obtained. Then the compatibility conditions of $v_i$ will lead to infinitely many local conservation laws of the SG equation \eqref{sgu}. It is noted that some trivial local conservation laws might be involved, and thus one needs to further exclude them.

Now let us first substitute the expansion  \eqref{vexpansion} into Eq. \eqref{bt1} with $\lambda$ replaced by $\delta$, and then vanish the coefficients of the same powers of $\delta$. From the zeroth order, we have $\sin(u/2-v_0/2)=0$, and thus
\begin{eqnarray}
v_0=u. \label{lsoluv0}
\end{eqnarray}
Making use of the result \eqref{lsoluv0}, the next five equations can be written as
\begin{eqnarray}
v_{1x}=-2 m \sin u,\label{localv1x}
\end{eqnarray}
\begin{eqnarray}
v_{2x}=-m v_1\cos u,\label{localv2x}
\end{eqnarray}
\begin{eqnarray}
v_{3x}=-m v_2\cos u+\frac 1 4 m v_1^2\sin u,\label{localv3x}
\end{eqnarray}
\begin{eqnarray}
v_{4x}=\frac 1 {24}m (v_1^3-24v_3)\cos u+\frac 1 2 m v_1v_2\sin u,\label{localv4x}
\end{eqnarray}
\begin{eqnarray}
v_{5x}=\frac 18m(v_1^2v_2-8v_4)\cos u-\frac 1{192}m(v_1^4-48v_2^2-96v_1v_3)\sin u.\label{localv5x}
\end{eqnarray}
Then the substitution of the expansion  \eqref{vexpansion} into Eq. \eqref{bt2} with $\lambda$ replaced by $\delta$, and the vanish of the coefficients of $\delta^i~(i=0,1,2,3,4)$ yield  a sequence of equations. The first equation also gives the solution \eqref{lsoluv0} and the next five equations have the results
\begin{eqnarray}
v_{1}=-\frac 2 m u_t,\label{localv1}
\end{eqnarray}
\begin{eqnarray}
v_{2}=-\frac 1 {m} v_{1t},\label{localv2}
\end{eqnarray}
\begin{eqnarray}
v_{3}=-\frac 1 m v_{2t}+\frac 1 {24} v_1^3,\label{localv3}
\end{eqnarray}
\begin{eqnarray}
v_{4}=-\frac 1 m v_{3t}+\frac 1 8 v_1^2v_2,\label{localv4}
\end{eqnarray}
\begin{eqnarray}
v_{5}=-\frac 1 m v_{4t}+\frac 1 8 v_1^2v_3+\frac 1 8 v_1v_2^2-\frac 1 {1920} v_1^5.\label{localv5}
\end{eqnarray}

Now, as before, it is ready to derive explicit local conservation laws from the consistent conditions 
$\partial_t v_{ix}=\partial_x\partial_t v_i~(i=1,2,3,4,5)$ for Eqs. \eqref{localv1x}-\eqref{localv5}. The first five local conserved densities read
\begin{eqnarray}
\rho_1=u_{xt}=m^2\sin u,
\end{eqnarray}
\begin{eqnarray}
\rho_2=u_{xtt}=m^2u_t\cos u,
\end{eqnarray}
\begin{eqnarray}
\rho_3=\frac {u_{tt}u_{xtt}} {u_t}=m^2u_{t t}\cos u,
\end{eqnarray}
\begin{eqnarray}
\rho_4=\frac {u_{xtt}u_{ttt}-u_t^2u_{tt}u_{xt}} {u_t}=-m^2u_tu_{tt}\sin u+m^2u_{ttt}\cos u,\label{lrho4}
\end{eqnarray}
\begin{eqnarray}
\rho_5&=&\frac {u_{xtt}u_t^2u_{tt}+2u_{xtt}u_{tttt}-2u_{xt}u_t^2u_{ttt}-u_{xt}u_{tt}^2u_t} {u_t}\nonumber\\
&&=-m^2(u_{tt}^2+2u_tu_{ttt})\sin u+m^2(2u_{t t t t}+u_t^2u_{t t})\cos u,\label{lrho5}
 \end{eqnarray}
and the corresponding local conserved flows are
\begin{eqnarray}
J_1=-u_{tt},
\end{eqnarray}
\begin{eqnarray}
J_2=-u_{ttt},
\end{eqnarray}
\begin{eqnarray}
J_3=-u_t^2u_{tt}-u_{tttt},
\end{eqnarray}
\begin{eqnarray}
J_4=-2u_tu_{tt}^2-u_{ttt}u_t^2-u_{ttttt},
\end{eqnarray}
\begin{eqnarray}
J_5=-16u_tu_{tt}u_{ttt}-3u_t^2u_{tttt}-u_{tt}u_t^4-5u_{tt}^3-2u_{tttttt}.
\end{eqnarray}
It is evident that the first two are trivial conservation laws, however, the others are all nontrivial.

\section{Summary and discussions}
In summary, by solving the linearized equations of the SG equation together its BT, not only three novel nonlocal symmetries, but also three special nonlocal conservation laws for the SG equation are obtained directly. Since these results depend on one parameter stemming from BT while not existing for the SG equation, therefore, the parameter expansion procedure is effective to generate three sequences of infinite number of nonlocal symmetries  and three sequences of infinite number of nonlocal conservation laws from the new special nonlocal symmetries and nonlocal conservation laws, respectively.  

Though it is difficult to derive the general formula for the sequence of infinitely many nonlocal symmetries and nonlocal conservation laws, it is quite obvious to write down their explicit expressions one by one. Here only one sequence of nonlocal symmetries and one family of nonlocal conservation laws are presented in detail. Furthermore, it is shown that infinitely many local and nonlocal conservation laws can also be generated via the same expansion procedure directly from BT of the SG equation. 

It is remarkable that after obtaining the nonlocal symmetries, one can go further to explore the symmetry reductions and even invariant solutions by means of the localization method. Besides, one can explore new integrable systems from new local and nonlocal conservation laws.

\section*{Acknowledgement}
The authors acknowledge the financial support by the National Natural Science Foundation of China (No. 11275123 and No. 11175092), Shanghai Knowledge Service Platform for Trustworthy Internet of Things (No. ZF1213), Talent Fund and K. C. Wong Magna Fund in Ningbo University.

\end{document}